\documentstyle[fleqn,12pt]{article}
\begin{document}
\title{Simple-minded estimate of the masses of baryons containing single
heavy quarks
\thanks{Supported in part by the KBN grants 2P302-112-06 and
2P302-076-07.}
}
\author{Agnieszka Zalewska\\ {\it Institute of Nuclear Physics, Krak\'ow,
Poland}
\\ Kacper Zalewski
\\
{\it Institute of Physics of the Jagellonian University}\\ {\it and}\\
{\it Institute of Nuclear Physics, Krak\'ow, Poland}}
\maketitle

\abstract{The masses of the yet undiscovered baryons containing single $c$ or
$b$ quarks are estimated from the known masses using the following rules: equal
distances in mass between the isomultiplets forming sextets, equal mass
differences between the corresponding spin one-half baryons containing $c$ and
$b$ quarks, hyperfine splittings inversely proportional to the masses of the
heavy quarks.}
\vspace{2cm}

According to the familiar SU(4) classification of baryons (cf e.g. \cite{PDG})
to every octet consisting of light ($d,\;u,\;s$) quarks only, corresponds, for
every heavy flavour $Q$, a sexted and an antitriplet of spin one-half baryons
containing each exactly one heavy quark $Q$. $Q$ denotes here $c$ or $b$,
because the $t$-quark decays, usually into a $W$ boson and a $b$-quark, before
it has time to hadronize. In each baryon from the sextet the light two-quark
system has spin one, which when combined with the spin one-half of the heavy
quark can produce besides the spin one-half baryon also a spin three-halves
baryon. This gives another sextet of baryons distinguished in the usual
notation by an asterisk. The usual notation (cf e.g. \cite{PDG}) is: $\Sigma_Q$
and $\Sigma_Q^*$ for the isotriplets from the sextets, $\Xi_{Q}'$ and
$\Xi_{Q}^*$ for the isodoublets from the sextets, $\Omega$ and $\Omega^*$ for
the isosinglets from the sextets, $\Lambda_Q$ for the isosinglets from the
antitriplets and $\Xi_Q$ for the isodoublets from the antitriplets. There is
some confusion about which spin one-half isodoublet should have the prime. We
adhere here to the convention, which leaves the prime for the yet undiscovered
isodoublet. One could object that $SU(4)$ symmetry is too badly broken to be a
useful guide, but the same predictions follow from $SU(3)$ applied to the light
diquark system. Altogether for $Q = c,b$ one expects 16 "ground state" baryons
(isomultiplets) with single heavy quarks.

These baryons are in the process of being discovered. The 1994 Particle Data
Group Tables \cite{PDG} quote $\Lambda_c(2.2851 \pm 0.0006)$ (here and in the
following the numbers in brackets following the symbol of the particle are its
mass and the corresponding error, both in GeV ), $\Xi_c(2.4677 \pm 0.0017)$
(for the isomultiplets we quote the average mass with the average error),
$\Sigma_c(2.4531 \pm 0.0007)$ and $\Lambda_b(5.641 \pm 0.050)$. More recent
results include $\Omega_c(2.7068 \pm 0.001)$ from the $WA89$ Collaboration
reported at the Moriond (1995) Conference \cite{OMC}, a result from SKAT
\cite{S*C} $\Sigma^*_c(2.530 \pm 0.007)$, the CLEO result $\Xi^*_c(2.643 \pm
0.002)$ and the results reported at the Brussels EPS (1995) Conference
\cite{BLO} $\Lambda_b(5.638 \pm 0.016)$ from ALEPH, $\Sigma_b(m_{\Lambda_b} +
0.173 \pm 0.009)$ (here and in the following the statistical and the systematic
error are added in quadrature) and $\Sigma^*_b(m_{\Lambda_b} + 0.229 \pm
0.009)$ from DELPHI.

In view of this progress on the experimental side, the theoretical activity,
which has been going on for more than 20 years (cf e.g. \cite{RGG}, for a
recent review cf. \cite{KKP}), has recently significantly increased. Thus e.g.
A. Martin and J. -M. Richard \cite{MRI} quote ten predictions for the mass of
the $\Omega_c$ ranging from $2.610$ GeV to $2.783$ GeV. A comparison with the
experimental result ($2.707$ GeV) shows that all these predictions are good
within $100$ MeV, which is remarkable in view of the variety of approaches
used. People have used potential models (cf e.g. \cite{MRI}) and
phenomenological fits inspired by such models (cf e.g. \cite{RDL},\cite{RLP}),
suitably modified MIT bag models (cf e.g. \cite{SAD}), suitably modified Skyrme
models (cf e.g. \cite{RRS}), lattice methods (cf e.g. \cite{BOW}) etc. This
strongly suggests that the baryon spectrum is to a large extent defined by its
general features common to most of the reasonable models. Finding the simple
rules behind the models is of some interest. One can see, whether a model is
really an improvement compared to the simple-minded version of the rules. One
can also immediately distinguish the expected from the unexpected, when a new
mass value is obtained from experiment.

In this note we try the following set of rules inspired by the heavy quark
effective theory.
\begin{itemize}

\item The mass difference between any spin one-half $b$-baryon and the
corresponding $c$-baryon is the same. Using the experimental masses of the
baryons $\Lambda_b$ and $\Lambda_c$ we find for this difference $ \delta_1 =
(3.353 \pm 0.016)$ GeV.

\item The isomultiplets in the sextets are equidistant in mass. In the light
decuplet the corresponding mass differences are $ m_{\Sigma^*} - m_\Delta  =
153$Mev, $m_{\Xi^*} - m_{\Sigma^*} = 148$ MeV and $m_\Omega - m_{\Xi^*} =
139$ MeV. We expect a similar scatter of the differences --- of a few MeV
around an average --- also for the sextets. Using the experimental masses of
$\Sigma_c$ and $\Omega_c$ we find for the mass difference between adjacent spin
one-half isomultiplets from the sextets $\delta_2 = 0.127$\footnote{Here and in
the following the experimental uncertainties of the input are ignored, when
they are much smaller than the other uncertainties.} GeV. An immediate
consequence of this assumption is that the known $\Xi_Q$ particles are members
of the antitriplets and not of the sextets.

\item The hyperfine splittings i.e. the mass differences between the members of
the spin one-half sextets and the corresponding members of the spin
three-halves sextets depend only on the mass of the heavy quark $Q$ and are
inversely proportional to this mass. The second part of this assumption is a
well-known leading term estimate from the heavy quark effective theory.
Physically, it follows from the observation that the hyperfine splitting is
proportional to the chromomagnetic moment of the heavy quark, which is
inversely proportional to the mass of this quark. The first part is more
model-dependent. For instance, Rosner \cite{ROS} quotes a model, where the
hyperfine splitting gets reduced by a factor of 0.84, when going from
$\Sigma_b$ to $\Xi'_b$ and by a further factor of 0.81 when going from $\Xi'_b$
to $\Omega_b$. We have chosen equal splittings in order to keep the intervals
between isomultiplets within the spin three-halves sextets equal. This choice
is also supported by the data for mesons, where the hyperfine splittings for
the $D$ and $D_s$ mesons are equal within a small experimental error. For $Q=c$
we find this splitting from the mass difference between the experimental mass
of the $\Xi^*_c$ and the interpolated mass of the $\Xi'_c$:

\begin{equation}
M_{\Xi'_c} = \frac{1}{2} \left( M_{\Sigma_c} + M_{\Omega_c} \right).
\end{equation}
This yields $\delta_{3c} = 0.063$ GeV for $Q = c$. For $Q=b$ we assume that the
splitting is reduced by the ratio $m_c/m_b$, as suggested by the heavy quark
approach. This factor can be estimated by a variety of methods, e.g. by
comparing the hyperfine splittings for the $B$ and $D$ mesons. The result is
about one third. Thus we put $\delta_{3b} = 0.021$ GeV.

\end{itemize}

On the whole, we use the experimental masses of six particles ($\Lambda_c,
\Lambda_b, \Sigma_c, \Xi_c, \Omega_c$ and $\Xi^*_c$) and the ratio
$\frac{m_c}{m_b}$ deduced from the meson data to fix our parameters. As a check
we can find three other particle masses, for which preliminary experimental
data is available:

\begin{eqnarray}
m_{\Sigma^*_c} &  =  & m_{\Sigma_c} + \delta_{3c} = (2.516 \pm 0.010)
\mbox{GeV},\\
m_{\Sigma_b} & = &  m_{\Sigma_c} + \delta_1 = m_{\Lambda_b} + (0.168
\pm 0.005)\mbox{GeV},\\
m_{\Sigma_b^*} & = & m_{\Sigma_b} + \delta_{3b} = m_{\Lambda_b} + (0.189 \pm
0.020)\mbox{ GeV}.
\end{eqnarray}

In order to eliminate the large uncertainty ($\pm 0.016$ GeV) in the parameter
$\delta_1$, we have introduced into the formulae for $Q = b$ the mass of the
baryon $\Lambda_b$. Ignoring the theoretical uncertainties given above, our
predictions deviate from the data by $-2.0,\;-0.6,\;-4.4$ standard deviations
respectively. Since the data is preliminary, it is probably premature to draw
conclusions from this comparison. Let us note, however, that if the
experimental indication that the hyperfine splittings for $Q=b$ and $Q=c$ are
similar were confirmed, this would be a serious difficulty for most present
models.

The theoretical uncertainties have been estimated (very crudely!) as follows.
The variations of hyperfine splittings quoted by Rosner \cite{ROS} would
increase the mass of $\Sigma_c^*$ and decrease the mass of $\Omega_c^*$ by
about $11$ MeV each. This is believed to be an overestimate \cite{ROS},
therefore, $10$ MeV was taken as a conservative estimate of this uncertainty.
The uncertainty for $\Sigma_b$ has been assumed to be comparable to the error
on the mass of the $B_s$ meson obtained from the formula $m_{B_s} = m_B +
m_{D_s} - m_D$. This is smaller than the experimental uncertainty of about $6$
MeV in $m_{B_s}$. Therefore, we guess an uncertainty of $5$ MeV. For the $Q=b$,
$J = \frac{3}{2}$ baryons, besides the usual uncertainty for $Q=b$ there is an
additional uncertainty due to the uncertainty of the assumption of hyperfine
splittings inversely proportional to the quark masses. This is large, since the
preliminary DELPHI measurement of the $\Sigma_b^*,\;\Sigma_b$ splitting gives
$56$ MeV, while we expect $21$ MeV. We guess the total uncertainty of our
prediction to be about $20$ MeV which, when combined with the experimental
error in quadrature, reduces the deviation for $\Sigma_b^*$ to about $1.8$
standard deviation.

For the yet undiscovered baryons our rules give:

\begin{eqnarray}
m_{\Xi'_c} & = & (2.580 \pm 0.005) \mbox{ GeV}\\
m_{\Omega_c^*} & = & m_{\Xi^*_c} + \delta_{3c} = (2.770 \pm 0.010)\mbox{ GeV}\\
m_{\Xi_b} & = & m_{\Xi_c} + \delta_1 = m_{\Lambda_b} + (0.183 \pm 0.005)
\mbox{GeV}\\
m_{\Xi'_b} & = & m_{\Sigma_c} + \delta_1 + \delta_2 = m_{\Lambda_b} + (0.295
\pm 0.010)\mbox{ GeV}\\
m_{\Omega_b} & = & m_{\Sigma_b} + 2\delta_2 = m_{\Lambda_b} + (0.422 \pm 0.010)
\mbox{ GeV}\\
m_{\Xi^*_b} & = & m_{\Sigma_b} + \delta_2 + \delta_{3b} = m_{\Lambda_b} +
(0.316 \pm 0.020)\mbox{ GeV}\\
m_{\Omega^*_b} & = & m_{\Xi^*_b} + \delta_2 = m_{\Lambda_b} + (0.443 \pm 0.020)
\mbox{GeV}
\end{eqnarray}

The uncertainty for $\Xi'_c$ was estimated from the errors in the analogous
predictions for the isomultiplets in the light decuplet: $m_{\Sigma^*} =
\frac{1}{2}(m_\Delta + m_{\Xi^*})$ and $m_{\Xi^*} = \frac{1}{2}(m_{\Sigma^*} +
m_\Omega)$. These errors are $-2$ MeV and $-5$ MeV respectively. Therefore, we
estimate the error on $m_{\Xi'_c}$ as $5$ MeV. For the mass of $\Xi_b$ the
uncertainty was assumed to be equal to that for the mass of $\Sigma_b$. For
the heavier baryons $\Xi'_b$ and $\Omega_b$, this uncertainty has been doubled.
The uncertainties for the remaining baryons have been discussed above.

It is instructive to compare our results with another purely phenomenological
(no explicit dynamics) set of predictions given by R. Roncaglia et al.
\cite{RLP}. For $Q = c$ there is exact agreement for $\Xi'_c$ and $\Omega^*_c$,
while for $\Sigma^*_c$ our prediction is lower by $4$ MeV. For $Q=b$, putting
in our formulae $m_{\Lambda_b} = 5.638$ GeV, one finds: exact agreement for
$\Omega_b$, a $\Xi_b$  heavier by $11$ MeV in our case and in all other cases
heavier baryons in the approach od Roncaglia et al. For $\Sigma^*_b,\;
\Xi^*_b,\; \Omega^*_b$ the differences are respectively $23$ MeV, $26$ MeV and
$9$ MeV. For $\Sigma_b$ and $\Xi'_b$ the differences are $14$ MeV and $17$ MeV.
The discrepancies are well within the stated errors of the two approaches.

Let us conclude with a few comments. A calculation of the baryon masses
with an uncertainty of $10$ MeV or less from
first principles would be much more interesting than the phenomenological fits.
For the moment, however, it is not yet in sight. The present approach is purely
phenomenological, but it is very simple and the physical assumptions are
transparent. When the experimental data for the heavy baryon masses become
available, it will be interesting to see, how it compares with the more
sophisticated approaches. The predictions are based on seven free parameters
taken from experiment: the masses of $\Lambda_c,\; \Xi_c,\;\Xi^*_c$ and
$\Lambda_b$ and the parameters $\delta_2,\;\delta_{3c},\;\delta_{3b}$ defined
in the text. The mass of $\Lambda_b$ may be traded for the parameter
$\delta_1$. This number of free parameters is not outrageous. A typical quark
model would have used four quark masses (assuming $m_u = m_d$) and three
parameters in the potential. The stated number of free parameters is somewhat a
matter of taste. All the parameters are constrained. Moreover, some parameters
may be presented as predictions of the theory. Suppose for example that our
assumption that $\delta_{3c} = 3\delta_{3b}$ works. It is inspired by (but not
rigorously derived from) the heavy quark approach and supported by the
experimental results for the hyperfine splittings in the mesonic sector,
therefore, one could present it as a theoretical result and reduce by one the
number of free parameters. In fact, assessing the number of free parameters in
various models one has to be very careful about such "hidden parameters". The
most risky assumption in our approach is the estimate of $\delta_{3b}$, which
contradicts the preliminary result for the mass of $\Sigma^*_b$. If this
experimental result is confirmed, one will have to increase $\delta_{3b}$
i.e. to shift up by equal amounts the expected masses of $\Sigma^*_b,\;
\Xi^*_b$ and $\Omega^*_b$. Then, however, a theoretical problem arises: why the
pattern of hyperfine energy splittings for heavy baryons is so different from
that for the mesons, or equivalently why the ratios of the hyperfine splittings
in baryons to those in mesons increases from below $0.5$ for $Q=c$ to about
$1.2$ for $Q=b$?  Let us repeat finally that our estimates of errors are very
crude, based on uncertain analogies --- though they are probably good
enough to distinguish the more uncertain from the less uncertain predictions.
To be sure, the theoretical errors quoted are understood as analogues of one
standard deviation and not as maximum conceivable errors.

\end{document}